# FA$_2$PbBr$_4$: synthesis, structure and unusual optical properties of two polymorphs of formamidinium-based layered (110) hybrid perovskite

Sergey A. Fateev [1#], Andrey A. Petrov [1#], Ekaterina I. Marchenko [1], Yan V. Zubavichus [2], Victor N. Khrustalev [3,4], Andrey V. Petrov [5], Sergey M. Aksenov[6], Eugene A. Goodilin [1,7], Alexey B. Tarasov *[1,7]

[1] Laboratory of New Materials for Solar Energetics, Faculty of Materials Science, Lomonosov Moscow State University; 1 Lenin Hills, 119991, Moscow, Russia

[2] Federal Research Center Boreskov Institute of Catalysis, Lavrentiev Ave. 5, 630090, Novosibirsk, Russia

[3] Inorganic Chemistry Department, Peoples' Friendship University of Russia (RUDN University), 6 Miklukho-Maklay Str., 117198, Moscow, Russia

[4] N.D. Zelinsky Institute of Organic Chemistry RAS, 47 Leninsky Prosp., 119991, Moscow, Russia

[5] Institute of Chemistry, Saint-Petersburg State University, 26 Universitetskii prospect, 198504, Saint-Petersburg, Russia

[6] Laboratory of Nature-Inspired Technologies and Environmental Safety of the Arctic, Kola Science Centre, Russian Academy of Sciences, 14 Fersman str., Apatity, 184209, Russia

[7] Department of Chemistry, Lomonosov Moscow State University; 1 Lenin Hills, 119991, Moscow, Russia

**ABSTRACT:** Small cations such as guanidinium and cesium can act as templating cations to form low dimensional phases (2D, 1D, 0D) in the case of excess of organic halides. However, such phases with the widely used formamidinium (FA$^+$) cation have not been reported so far. In this study, we discovered two novel low dimensional phases with composition of FA$_2$PbBr$_4$ and investigated the prerequisites of their formation upon crystallization of FABr-excessive solutions of FAPbBr$_3$. We found that both phases have the structure of (110) layered perovskite but is represented by two different polymorphs with "eclipsed" and "staggered" arrangement of adjacent layers. It was shown that FA$_2$PbBr$_4$ phases usually exist in a labile equilibrium with FAPbBr$_3$ 3D perovskite and can form composites with it. The optical properties of both polymorphs were comprehensively studied by means of absorption spectroscopy, diffuse reflection spectroscopy and photoluminescence spectroscopy. DFT calculations were applied to investigate the band structure of the FA$_2$PbBr$_4$ and to corroborate the conclusions on their optoelectronic properties. As a result, we found that FA$_2$PbBr$_4$ phases irradiated by UV can exhibit effective green photoluminescence due to a transfer of excitation energy to defective states or 3D perovskite inclusions.

## 1. Introduction

The outstanding optical and electronic properties of organo-inorganic haloplumbates with a perovskite structure provided an unprecedented surge of interest as a new generation of semiconductor materials for efficient photovoltaic and optoelectronic devices. One of the most important features of this innovative class of semiconductors is the extraordinary tunability of their optoelectronic parameters through adjusting the connectivity of the inorganic anionic framework [1].

The majority of structural design strategies are based on the concept of achieving the desired electronic dimensionality of a material by controlling the dimensionality of its inorganic framework[2], which is commonly reduced to layered (2D) or lower dimensional (1D, 0D) structures by introducing bulk organic cations acting as a template for the structure. A selection of a templating cation to form a framework of the desired topology is therefore a key issue for rational structural design of such materials.

Depending on the size and shape, an organic cation can either predetermine the only possible topology of the inorganic framework or provide several variants of co-packing with polyiodoplumbate anions. The first case is observed mainly for large cations with a long carbon chain, resulting in a set of structures practically limited by the class of (100) perovskites[3]. The second case is realized for relatively small cations with a size comparable to the Pb–X bond length (~3 Å).

These small cations (A$^+$) can form series of compounds with different ratio R = [A$^+$]/[PbX$^+$] and different packaging. An increase of R should imply a decrease in the number of halogen atoms in the first coordination sphere of the A$^+$ and a simultaneous increase in the number of contacts between A$^+$ cations. This can be well illustrated by a series of compounds with guanidinium (Gu$^+$) cations GuPbI$_3$ (1D)[4], Gu$_2$PbI$_4$ (2D)[5], Gu$_3$PbI$_5$ (1D) [6], Gu$_4$PbI$_6$ (0D)[6] and with cesium cations CsPb$_2$Br$_5$ (2D)[7], CsPbBr$_3$ (3D)[8], Cs$_4$PbBr$_6$ (0D)[8]. For such systems (CsBr-PbBr$_2$ and GuBr-PbBr$_2$), stoichiometry of the precursors strongly affects the structural dimensionality of the forming phase. Surprisingly, so far there have been no attempts to isolate any low-dimensional phases with R > 1 ("Pb-deficient") with the most frequently used formamidinium (FA$^+$) and methylammonium (MA$^+$) cations, while the size of these cations lies exactly between Gu$^+$ and Cs$^+$. This fact motivated us for the current research that yielded unexpected results presented in this article.

In this study, we investigated the possibility of low dimensional phases formation from the solutions of FABr-PbBr$_2$ and MABr-PbBr$_2$. We showed that, in the case of an excess of organic halides (FABr, MABr), two novel phases of (110) perovskite with composition FA$_2$PbBr$_4$ can be obtained in pure form whereas MA$_2$PbBr$_4$ does not exist. We determined the crystal structure of the two observed polymorphs of FA$_2$PbBr$_4$ and characterized their optoelectronic properties using absorption spectroscopy, diffuse reflection spectroscopy (DRS), photoluminescence excitation (PLE) and steady-state photoluminescence (PL) spectroscopy. Surprisingly, these two layered phases exhibit drastically different optical properties: while the trigonal modification has a noticeable photoluminescence, the monoclinic phase is only capable of transmitting excitation to adjacent 3D perovskite. DFT calculations were preformed to investigate the band structure of the FA$_2$PbBr$_4$ and to analyze the origins of its peculiar photoluminescent properties.

## 2. Experimental section

**Materials and methods.** Formamidinium bromide (CH(NH$_2$)$_2$Br = FABr, Dyesol), methylammonium bromide (CH$_3$NH$_3$Br = MABr, Dyesol), lead bromide (PbBr$_2$, 99.99%, TCI), dimethylsulfoxide (DMSO, anhydrous, > 99.9%, Sigma-Aldrich), dimethylformamide (DMF, anhydrous, > 99.8%, Sigma-Aldrich), chlorobenzene (>99,8%, Sigma-Aldrich) and other antisolvents were commercially purchased.

All solutions were prepared at room temperature under stirring during 2 h in argon filled glove box, unless otherwise indicated.

**Single crystal growth.** Experiments to isolate low dimensional $FA^+$ and $MA^+$ containing phases were made using several solutions in DMF with $FABr/PbBr_2$ ratio from 2 to 3 and the same solutions for the case of MABr excess.

Then we placed drops of all solutions on quartz glass and observed the process of crystal growth. Typical crystals from each droplet were isolated from the solution using a nylon loop, blotted and immediately transferred for X-ray measurements and solved as described below.

**Single Crystal X-ray Diffraction.** The single-crystal X-ray diffraction data set were collected on the 'Belok' beamline of the National Research Center 'Kurchatov Institute' (Moscow, Russian Federation) using a Rayonix SX165 CCD detector. In total, 720 frames were collected with an oscillation range of 1.0° in the φ scanning mode using two different orientations for the crystal. The semi-empirical correction for absorption was applied using the Scala program[9]. The data were indexed and integrated using the utility iMOSFLM from the CCP4 software suite[10]. For details, see Table S1. The structure was solved by intrinsic phasing modification of direct methods and refined by a full-matrix least-squares technique on F2 with anisotropic displacement parameters for all non-hydrogen atoms. The hydrogen atoms were objectively localized from the different Fourier-maps and refined within the riding model with fixed isotropic displacement parameters [$U_{so}(H) = 1.2U_{eq}(N,C)$]. All calculations were carried out using the SHELXTL program suite[11].

**Thin films preparation.** Glass substrates were cleaned with detergent, flushed with distilled water and then sequentially washed in ultrasonic baths in acetone, isopropyl alcohol and distilled water. Substrates were further cleaned with UV-ozone for 15 min prior to their use.

A homogeneous solution of $FABr:PbBr_2$ = 2:1 in DMSO-CB (1:1) mixture was prepared by slow drop-wise adding of 0.5 ml of chlorobenzene to equal volume of a 2.2 M $FA_2PbBr_4$ solution in DMSO at 110°C under vigorous stirring (to reach the concentration of resulting solution of 1.1M). Then the solution was cooled down to 60 °C and used for spin-coating of phase-pure $m-FA_2PbBr_4$ films.

The optimal parameters for obtaining a thin film of $t-FA_2PbBr_4$ were determined as the following: 1.25M solution of $FABr:PbBr_2$ = 2.15:1 in DMSO, rotation rate of 4000 rpm for 50 s with CB dropwise added after 25 s and annealing at 110 °C for 10 min followed by annealing at 90 °C for 10 min. It is important to note that one of the key conditions of single-phase film fabrication, as well as its further storage, is low humidity, therefore, all operations must be carried out in a glove box.

**Optical measurements.** The absorption spectra of the films were recorded both in transmission mode (on a Jasco V-550 UV/Vis spectrophotometer in the wavelength range of 300–700 nm at a scanning speed of 100 nm/min) and in the diffuse reflectance mode (on a Perkin Elmer Lambda 950 spectrophotometer in the wavelength range of 300–800 nm) with wavelength resolution 1.0 nm. Steady-state PL measurements were performed on a microscope assembled using Thorlabs optomechanical components. Samples were photoexcited using a 405 nm laser (InTop, Russia) delivering 30 ps (FWHM) pulses in the continuous wave mode. A Flame (Ocean Optics, UK) spectrometer was used to record PL spectra.

**Powder X-ray diffraction** analysis was provided using Rigaku D/MAX 2500 (Japan) with a rotating copper anode (CuKα irradiation, 0.15418 nm) in the Bragg–Brentano geometry. XRD patterns were sequentially measured at a scanning speed of 0.2 deg/s.

**Crystal chemical analysis.** The fragmentation of the unit cell volume into "partial" specific volumes of moieties of the interest ($FA^+$ cations, [$PbBr_4$]) in the structure is performed using Voronoi-Dirichlet polyhedra (VDP). The main algorithms of constructing VDP are implemented in ToposPro program package.[12] VDP can often characterize not only relative, but also absolute atomic size in the crystal structure. [13]

To determine the degree of distortion of $PbX_6$ octahedra in the structure, we used the equation introduced in ref. [14]:

$$\Delta d = \frac{1}{6}\sum \left[\frac{d_n - d}{d}\right]^2, \qquad (1)$$

where $d_n$ - individual B-X distances, d - arithmetic mean values of the individual B-X distances.

**DFT calculations.** For calculation of electronic structure of $FA_2PbBr_4$ we used the DMol3 module from Materials Studio software package[15,16]. In this program, the DFT method is realized on the basis of atomic functions. We applied the DNP+ basis (Double Numerical plus polarization with addition of diffuse functions) and spin-orbital coupling.

## 3. Results and Discussion

The crystallization of colorless crystals of $FA_2PbBr_4$ was observed for DMF solutions with $FABr/PbBr_2$ =2-3. Depending on the conditions, two crystallization paths can be observed: in the case slow evaporation of a large unperturbed drop, thin plate-like crystals grow on the surface of the solution (1), whereas in the case of relatively rapid evaporation (small droplets), faceted prism-shaped crystals grow in the volume of the solution (2) (Figure S2). We resolved the structure of both types of crystals and found that they are two different phases with the same composition – $FA_2PbBr_4$. (Table 1). From a crystallographic point of view, these two phases are polymorphs, one of which is triclinic and the other is monoclinic (for simplicity, we denote them further as $t-FA_2PbBr_4$ and $m-FA_2PbBr_4$ respectively).

Surprisingly, these two polymorphs crystallize sometimes simultaneously from a single droplet of solution (Figure S3) This implies that crystallization is controlled more by kinetics of solvent evaporation than by thermodynamics and indicates that the lattice energies of two polymorphs are close. Under equilibrium conditions without evaporation of the solvent, the formed crystals of the $t-FA_2PbBr_4$ coexist with a saturated solution for an unlimited time, but stirring triggers its slow recrystallization into $m-FA_2PbBr_4$. Also, phase-pure $m-FA_2PbBr_4$ can be obtained via CB vapour diffusion.

In the same manner, we attempted to obtain crystals of MABr-excessive phases from solutions of $MABr:PbBr_2$ with different precursor ratios (up to 3:1). However, in all the cases, only bright orange $MAPbBr_3$ perovskite crystals grew from DMF and DMSO solutions as well known for this system [17].

### Peculiarities of the $FA_2PbBr_4$ crystal structure

Both $FA_2PbBr_4$ polymorphs have crystal structure of (110)-oriented layered perovskite, which consists of corner-connected $PbBr_6$ octahedra forming corrugated layers separated with formamidinium cations (Figure 1, Table 1).

Table 1. Crystallographic Data for two polytypes of $FA_2PbBr_4$.

| Polytype | $t-FA_2PbBr_4$ | $m-FA_2PbBr_4$ |
|---|---|---|
| Appearance | Large thin plate-like crystals | Faceted prism-shaped crystals |

| Crystal system | Triclinic | Monoclinic |
|---|---|---|
| Space group | $P\bar{1}$ | $P2_1/c$ |
| Unit cell parameters | a = 8.5642(17) Å; b = 11.884(2) Å; c = 14.312(3) Å<br>α = 65.92°; β = 73.01°; γ = 83.76° | a = 13.1313(9) Å; b = 7.9763(6) Å; c = 11.7972(8) Å<br>α = 90.0°; β = 91.52°; γ = 90.0° |
| Volume | 1271.7(6) Å³ | 1271.7(6) Å³ |
| Z | 2 | 4 |
| Density (calculated) | 3.222 g/cm³ | 3.318 g/cm³ |

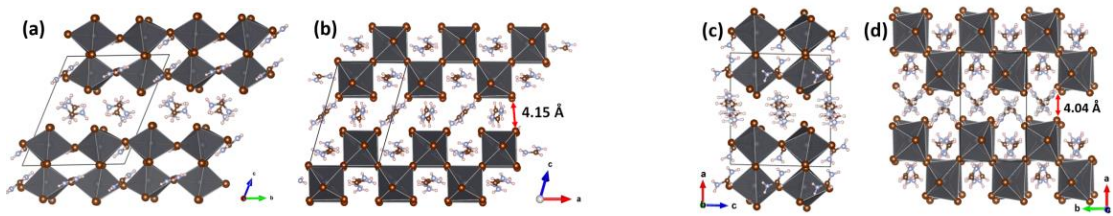

Figure 1. Projection of t-$FA_2PbBr_4$ crystal structure on (a) 0kl and (b) h0k plane and m-$FA_2PbBr_4$ on (c) h0k and (d) hk0 planes.

The main distinction between the two polytypes is different stacking of adjacent inorganic layers which can be described in terms of the layer shift factor (LSF)[18]. The m-$FA_2PbBr_4$ (Figure 1, b) has a staggered[3] stacking of layers (LSF = [0.5, 0.06]) typical for the majority of previously reported (110) layered bromide perovskites [5,18–26], while the t-$FA_2PbBr_4$ (Figure 1, a) represents the first known structure of a bromoplumbate with an eclipsed stacking of the perovskite layers (LSF = [0.06, 0.09]).

Both structures are stabilized by hydrogen bonds formed by $FA^+$ cations, each connected with several neighboring bromide anions thus templating the corrugated inorganic framework of $FA_2PbBr_4$. In both polymorphs formamidinium cations simultaneously occupy two different types of positions – intralayer and interlayer (see detailed analysis of $FA^+$ positions in Figure S4) In terms of size, the former virtually corresponds to the cuboctahedral void occupied by the $FA^+$ cation in $FAPbBr_3$ (3D) perovskite. The only difference is the absence of one $Br^-$ in the first coordination sphere. In contrast, $FA^+$ cation in interlayer positions forms hydrogen bonds simultaneously with two adjacent layers, thereby binding them together. Interestingly, both for the cation in the interlayer positions and in the interlayer ones, the coordination number is equal to 12, but the anions in the nearest environment are partially replaced by cations ($FA^+$). An average specific volume of $FA^+$ cations in different positions remains the same and has a negligible difference compared to $FAPbBr_3$ (Figure S13, Table S5).

The second important structural feature of two $FA_2PbBr_4$ polymorphs is a very short interlayer distance. Due to a small size of the $FA^+$, the closest Br···Br distances between the layers in the t-$FA_2PbBr_4$ and m-$FA_2PbBr_4$ are 4.12 Å and 4.04 Å, respectively. It is even smaller than the edge length of the $PbBr_6$ octahedron. To the best of our knowledge, this is one of the smallest interlayer distance reported for bromide perovskites so far (Table S7).

The unique feature of the t-$FA_2PbBr_4$ is the lowest degree of inorganic lattice distortion (Δd) among all the (110) bromoplumbate perovskites known so far. This means that all the Pb–Br bonds have approximately same length. Indeed, in the most of hybrid halide perovskites the inorganic framework distortion occurs primarily because the organic cation (or its part bearing the positive charge and interacting with inorganic lattice) mismatches the size of the cuboctahedral void. On the contrary, the $FA^+$ cation is only slightly larger than the cuboctahedral void (tolerance factor about 1.08)[27] and, as a result, it keeps the structure slightly distorted only.

An important feature of the t-$FA_2PbBr_4$ polymorph is the presence of a significant number of defects and stacking faults in single crystals, which is reflected in the rather low quality of the collected XRD data and high R-factor.

### Comparison of "templating ability" of $FA^+$ and $MA^+$ cations

In order to understand why the layered perovskite is formed in the case of formamidinium, but does not exist for methylammonium, we compared the structures of the known (110) lead halide perovskites with various organic cations (listed in Table S7) and revealed the common features of the organic cations. We propose three main requirements that an organic cation has to meet in order to template the (110) layered structure: it should be able to form multiple hydrogen bonds and at the same time be rigid and "linkering" (i. e. be connected with adjacent perovskite layers). Another common, but not necessary prerequisite is the planar geometry of the cations, which determines their effective packing. Simultaneous compliance with these requirements can be inherent in both singly and doubly charged cations making them effective templating agents. Generally speaking, this explains why only a few (110) perovskites are known so far (Table S7).

Obviously, the formamidinium satisfies all these three requirements, whereas methylammonium does not. As a result, a similar (110) layered perovskite structure does not form with $MA^+$ cations.

### $FA_2PbBr_4$ films deposition features: overcoming competing crystallization of phases

To investigate further the properties of $FA_2PbBr_4$ phases we obtained it in a form of thin films. However, the deposition of phase-pure films is hindered by competing crystallization of 3D perovskite phase due to a labile equilibrium (1) between these solid phases:

$FA_2PbBr_4^{(s)} \leftrightarrow FAPbBr_3^{(s)} + FABr$     (1)

As a result, in the case of solutions with $FABr:PbBr_2 = 2:1$ both in DMF and DMSO with or without of antisolvents an admixture of the 3D perovskite phase has always been observed in the films obtained by conventional spin-coating. To overcome this issue, we proposed two approaches.

The first is a simple addition of 15% FABr excess in solution to shift the equilibrium (1) to the right which results in single phase films of t-$FA_2PbBr_4$ (Figure 2). The second one is a new technique of film processing from a homogeneous solvent – antisolvent mixture (DMSO-CB), prepared at elevated temperature. The essence of latter approach is to reach slowly the supersaturated state of the solution before spin-coating, in which nuclei of the composition corresponding to the composition of the solution form predominantly. The films obtained in this way are characterized by high transparency and uniform pale fawn color. SEM images show that the films are uniform and have no pinholes (Figure S7 c). The thickness of the film is about 0.85 μm (Figure S7 b). The crystallites have elongated form with average length about 1.5 μm (up to 3 μm) and width about 0.5 μm (Figure S7 a).

It should be noted that, being a layered perovskite, both polymorphs of the $FA_2PbBr_4$ are very sensitive to water vapors. Even at relative humidity of 30-40% they decompose forming $FAPbBr_3$. We associate this with the ability of water to solvate preferably and "to extract" FABr efficiently as a protic solvent[28], shifting the equilibrium (1) towards the 3D perovskite (Figure S5).

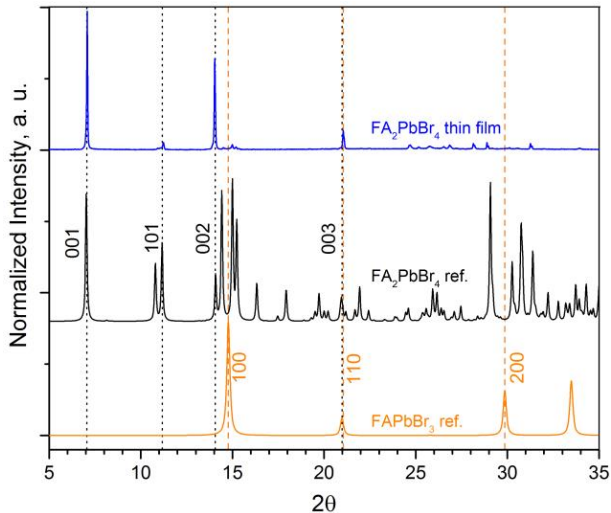

Figure 2. XRD pattern of phase-pure t-$FA_2PbBr_4$ thin film and simulated powder diffraction patterns of t-$FA_2PbBr_4$ and $FAPbBr_3$.

In addition, we studied the effect of the partial replacement of $FA^+$ cations with $MA^+$ in solutions on the phase composition of the films. In particular, it was found that at a $[FA^+]/[MA^+]$ ratio R = 0.75 (the overall ratio of organic cations to $PbBr_2$ was kept equal to 2), the $FAPbBr_3$ phase is formed predominantly along with $FA_2PbBr_4$. With a decrease of R, the amount of 3D phase of perovskite increases even more and at R = 0.25 XRD does not show any traces of layered perovskite (Figure S6). Simultaneously, the 100 and 200 reflections of $FAPbBr_3$ shift to larger angles, indicating the formation of $(FA_xMA_{1-x})PbBr_3$ solid solutions.

Thus, it was shown that $FA_2PbBr_4$ does not form substitution solid solutions with methylammonium. On the contrary, introducing $MA^+$ results in destabilization of the layered phase.

### Optical properties and electronic structure

A diffuse reflectance spectra of the t-$FA_2PbBr_4$ and m-$FA_2PbBr_4$ thin film on quartz glass reveal an absorption edge at 455 and 420 nm respectively. According to the analysis of the spectra by the Tauc plot the optical band gaps of t-$FA_2PbBr_4$ and m-$FA_2PbBr_4$ are 2.82 eV and 3.00 eV, respectively (Figure 3). The former is the lowest value of optical $E_g$ known for (110) bromide perovskites (Table S8). As shown below, we associate such a small value of $E_g$ with weak orbital overlapping (Br…Br) due to very short interlayer distance compared to other (110) bromide perovskites (4.15 Å for t-$FA_2PbBr_4$, see Table S7).

An absorption spectrum of the m-$FA_2PbBr_4$ recorded in transmission mode exhibits a clear excitonic band centered at 3.15 eV (Figure S8). The flat maximum of the exciton band most likely indicates the presence of several energy-related exciton transitions associated with anisotropic deformation of the octahedra[29]. Fitting of m-$FA_2PbBr_4$ spectrum according to the Elliott model gives a bandgap ($E_g$) of continuum transitions at 3.3 eV and an exciton binding energy of 170 meV, which are very typical values for bromoplumbate perovskites.

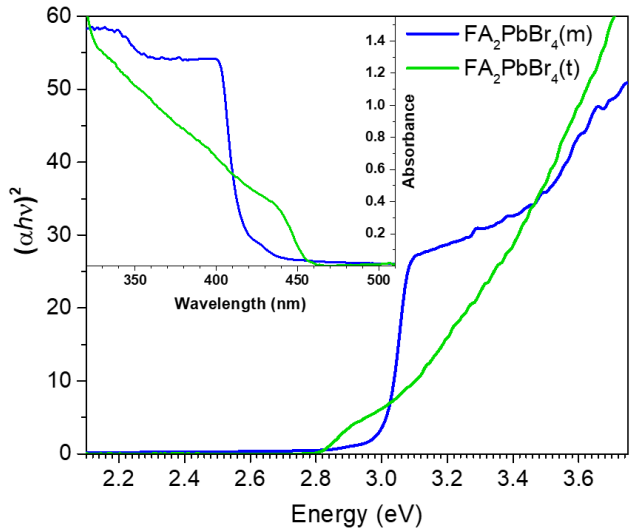

Figure 3. The diffuse reflectance spectra of the triclinic (t) and monoclinic (m) polytypes of $FA_2PbBr_4$ in Tauc coordinates and in Absorbance units (in the inset).

To reveal the origin of electronic transitions in t-$FA_2PbBr_4$, we performed DFT calculations using the refined crystallographic data (see details in SI). The calculations reveal a direct bandgap of 3.14 eV (Figure 5a, Table S2), which fairly agrees with the experimental value. The top of the valence band formed by p-orbitals of Br and bottom of the conduction band formed by p-orbitals of Pb (Figure 4b), which is typical for hybrid lead bromide perovskites.[25] Therefore, near-bandgap electronic transitions is dominated by 6p $Br^-$ to 6p $Pb^{2+}$ states. A relatively small dispersion of the bands along the layers directions (x,y) in $FA_2PbBr_4$ compared to (100) layered bromide perovskites is explained by a lower electronic dimensionality of the "corrugated" (110) layered structures[2,25].

The bandgap of the t-$FA_2PbBr_4$ is noticeably smaller than that for the most of related (110) lead bromide perovskites, which can be associated with both an unusually small interlayer distance (d) and a relatively small distortion of inorganic framework (Table S7). To distinguish these two effects, we performed calculations of the electronic structure for two series of model structures (Fig.S9, S10) derived from the experimentally solved structure of the $FA_2PbBr_4$ (see detailed description in SI). The structures of the first series preserve the geometry of inorganic framework of the real structure (the octahedra distortion and tilting) and have different $d_{mean}$ from 3.16 to 9.65 Å (Table S3, Figure S9). Expectedly, the main change in the band structure is an increase of the bandgap with the increase of the interlayer distance (Figure 4c). Moreover, $E_g(d)$ rises mainly within the range of 3.5 - 4.5 Å and reaches further a plateau at d > 4.6 Å (Fig. 6c). Such a dependency of $E_g(d)$ originates obviously from the contribution of a relatively weak interlayer electronic coupling due to partial overlapping of the orbitals of terminal Br atoms. Interestingly, this Br⋯Br overlapping contributes in the band structure even at the distances significantly exceeding the doubled value of the van der Waals radius of Br (1.85 Å) and doubled ionic radius the bromide (1.96 Å), which are usually considered as the upper limits for weak orbital overlapping[29]. For another series of model structures with undistorted inorganic framework and a similar set of interlayer distances (Table S9, Fig.S10), a $E_g(d)$ curve has a similar shape, although it is downshifted by a constant of -0.11 eV (Fig. 6c). Based on these two tendencies, we evaluated quantitatively the contribution of two main factors to the total bangap value: -0.045 eV for the weak interlayer orbital overlapping and +0.11 eV for the octahedra distortion.

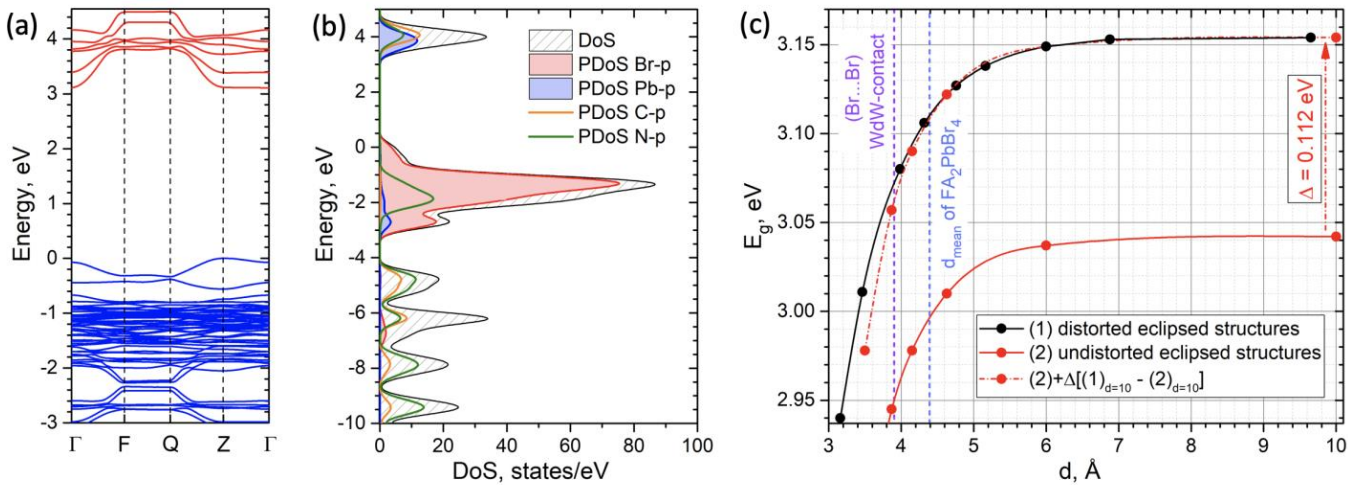

Figure 4. Band structure (a) and electronic density of states (DOS) (b) of $FA_2PbBr_4$(t) obtained by DFT calculations. The dependences (c) of bandgap values ($E_g$) from the distance between inorganic layers (d) for two series of model structures.

The photoluminescent properties of the two polymorphs differ dramatically. Whereas phase-pure m-$FA_2PbBr_4$ exhibits no emission at room temperature, a thin film of t-$FA_2PbBr_4$ shows intense green photoluminescence (PL) with a maximum centered at 2.37 eV (523 nm), despite a negligible absorption in the visible range (Figure 5 c). The observed emission band for the films is far from that for 3D perovskite ($FAPbBr_3$), and, in addition, is noticeably wider. Therefore, it is hardly possible to attribute the observed emission directly to the perovskite phase.

To further reveal a nature of the radiative recombination in the t-$FA_2PbBr_4$ we collected the spectra from single crystals. An important feature of t-$FA_2PbBr_4$ crystals is the presence of colorless and pale-yellow areas (Figure 5 a). We recorded the emission spectra from these two areas using a fluorescent microscope under the excitation by diode laser at 405 nm. The spots in transparent areas ("1" in the Figure 5) demonstrate a main narrow peak at 2.91 eV and a second broad peak of lower intensity centered approximately at 2.46 eV. The first peak can be clearly attributed to intrinsic exciton radiative recombination. For the spots in the pale yellow regions of the crystal, an inverse intensity ratio of two peaks is observed.

The origin of the broad band at 2.46 eV with a significant Stocks shift and FWHM of 0.2 eV is rather unclear. The most obvious explanation could be the presence of an impurity phase with a smaller band gap. However, this band is far from a position of PL maximum characteristic for $FAPbBr_3$ thin films (2.23 eV) [30] and also does not match to emission peak of $FAPbBr_3$ seed crystals (2.16 eV) formed in FABr-excessive solution in DMF (Figure 5d, "4"). Moreover, partially colored and completely transparent crystals are indistinguishable using X-ray methods (neither powder-XRD nor single-crystal diffraction). Therefore, it can be concluded that the coloring is not associated with the presence of a new phase, but rather is a consequence of the appearance of some centers of radiative recombination inside the phase.

Such centers can correspond either to self-trapped excitons (STEs) or to a special type of bulk or surface static defects. The STEs are widely known for halide perovskites and especially for (110) lead bromide perovskites. In most cases recombination via STEs are associated with a large Stocks shift and broad FWHM. However, the width of t-$FA_2PbBr_4$ second emission band is significantly narrower in comparison with a typical values for related perovskites (Table S8).

In addition, as can be seen from Figure 5 d, the spectrum taken from the colored area of the t-$FA_2PbBr_4$ crystal is very similar to the spectrum from a thin film of the same composition. The simultaneous increase of the intensity of the band at 2.46 eV and disappearance of an excitonic band for a thin film compared to single crystal implies the dependence of PL from the surface-bulk ratio. Accordingly, the observed emission band should rather be attributed to surface defects.

The latter assumption was further corroborated by means of photoluminescence excitation spectroscopy (PLE). The PLE spectrum was measured from the same thin film and showed an excitonic absorption edge near 3 eV and an additional band between 2.8 eV and 3.0 eV. This band matches well with the observed pale yellow color characteristic for thin films of t-$FA_2PbBr_4$. Taking into account the luminescence spectra, this band can be interpreted as below bandgap absorption caused by surface defects. Additionally, the shape of the PLE spectrum implies the effective energy transfer from the intrinsic excited states of t-$FA_2PbBr_4$ to the (defect) states related to the radiative recombination.

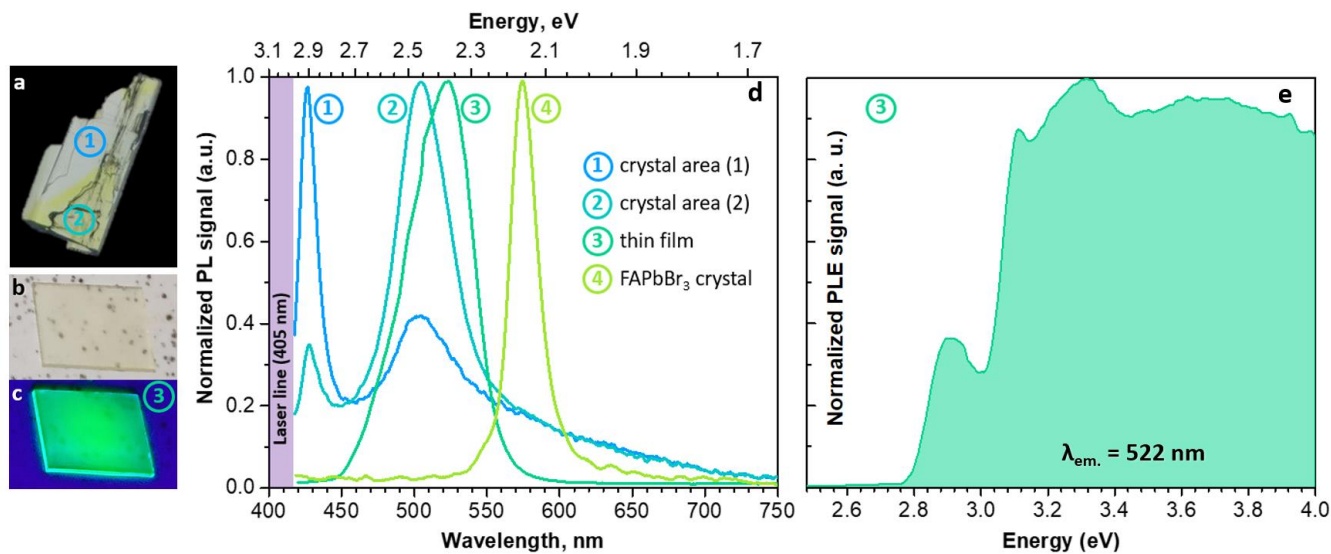

Figure 5. The appearance of $FA_2PbBr_4$(t) single crystal (a), from two areas (1 and 2) of which the PL spectra were recorded. The thin film of $FA_2PbBr_4$(t) on glass under visible (b) and UV (c) irradiation and corresponding PL (d, 3) and PLE spectra (e). All spectra on figure (d) were obtained under irradiation of diode laser emitting at 405 nm. Spectrum (4) from $FAPbBr_3$ perovskite crystals was taken under similar conditions and presented for comparison.

The origin of the assumed defect states may be related to multiple stacking faults in the structure of t-$FA_2PbBr_4$, as well as to similar intergrowth structures between the polymorphs. Another explanation relates to the nanoparticles of $FAPbBr_3$ which may exist in the film in a small concentration remaining undetectable by the applied methods of composition characterization. According to the literature, $FAPbBr_3$ nanoparticles with a size smaller than 10 nm demonstrate a blue shift of PL about 0.1 eV [31,32] compared to a thin film (2.2-2.3 eV). This implies that only very small nanoparticles will have enough blue shift to explain the observed emission peak for the films. Noteworthy, the narrow bright emission induced by lower-bandgap 3D perovskite nanoparticles "embedded" inside the matrix of larger-bandgap material is a known phenomenon for analogous cases of $CsPbBr_3$ and $Cs_4PbBr_6$ respectively [33,34]. Therefore, the presence of embedded $FAPbBr_3$ nanoparticles in a small concentration may be plausible explanation of photoluminescence properties of t-$FA_2PbBr_4$ films.

### 4. Conclusions

To sum up, we investigated the formation of low dimensional phases from FABr-excessive $FABr-PbBr_2$ solutions and discovered new layered (110) bromoplumbate perovskite $FA_2PbBr_4$ represented by two polymorphs – triclinic and monoclinic. The first polymorph demonstrates a unique eclipsed manner of the perovskite layers stacking whereas the monoclinic one has a staggered one typical for (110) perovskites. Both structures are characterized by unusually small interlayer distance of 4.15 Å for t-$FA_2PbBr_4$ and 4.04 Å for m-$FA_2PbBr_4$. In addition, t-$FA_2PbBr_4$ has the most undistorted inorganic framework among all (110) bromoplumbate perovskites known so far.

The optoelectronic properties of $FA_2PbBr_4$ polymorphs were comprehensively characterized from both experimental and theoretical points of view. It was found that t-$FA_2PbBr_4$ has an optical bandgap of 2.8 eV, which is the lowest value among (110) bromide perovskites, whereas a bandgap of m-$FA_2PbBr_4$ is 3.0 eV. The relatively "narrow" bandgap of t-$FA_2PbBr_4$ was explained via modelling of band structure for series of related structures. Based on model calculations, we evaluated quantitatively the contribution of the overall distortion of inorganic framework and the weak interlayer orbital overlapping to the total bandgap value of $FA_2PbBr_4$ as -0.045 eV and +0.11 eV respectively.

We found that single crystals of t-$FA_2PbBr_4$ demonstrate simultaneously two photoluminescence bands: free excitonic one at 2.91 eV and another band at 2.46 eV with a broad FWHM. The second band is associated with radiative recombination via defect states existing on the surface of the $FA_2PbBr_4$ single crystals. According to PLE these defects are responsible for intense green photoluminescence of phase-pure t-$FA_2PbBr_4$ thin films with a maximum at 2.37 eV.

## ASSOCIATED CONTENT

The Supporting Information is available free of charge via the Internet at http://pubs.acs.org.

## AUTHOR INFORMATION

**Corresponding Author**

*alexey.bor.tarasov@yandex.ru

**Author Contributions**

#These authors contributed equally.

The manuscript was written through contributions of all authors. All authors have given approval to the final version of the manuscript.

**Notes**

The authors declare no competing financial interest.

**Funding Sources**

This research was financially supported by the Russian Science Foundation (Project No. 19-73-30022).

ACKNOWLEDGMENT

The authors are grateful to Aleksandra Shatilova for her help with experiments and to Artem Ordinartsev for introducing band structure analysis through python scripting.

DFT calculations were carried out using computational resources provided by Resource Center "Computer Center of SPbU".

The X-ray powder diffraction analysis was provided using Rigaku D/MAX 2500 diffractomer (the Faculty of Materials Science, Lomonosov Moscow State University).

**TOC:**

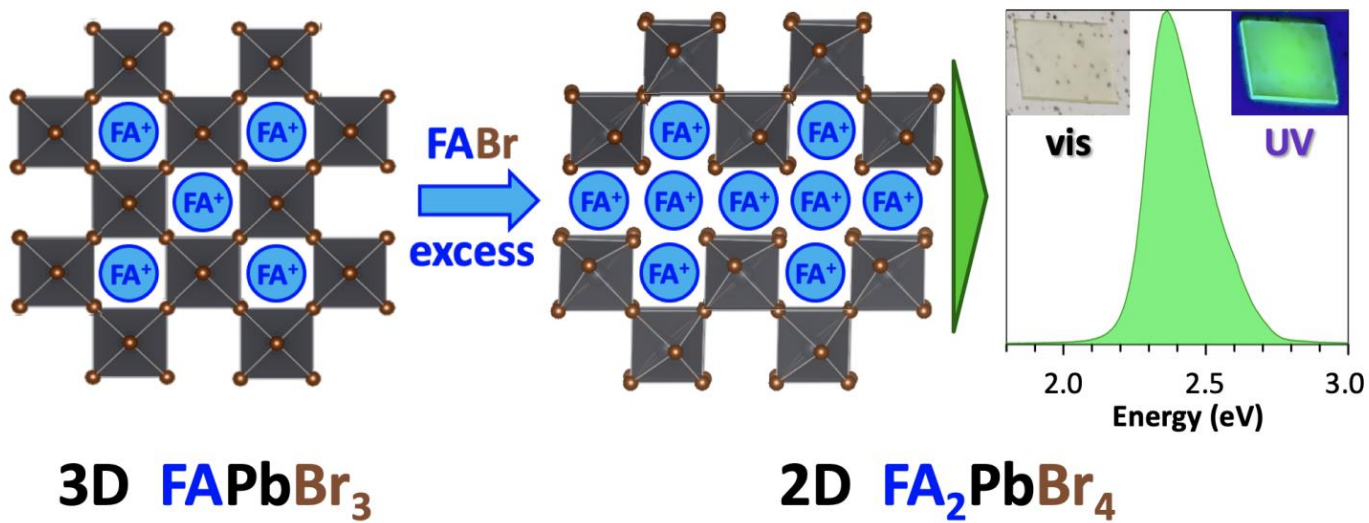

# Supporting Information

## FA$_2$PbBr$_4$: synthesis, structure and unusual optical properties of two polymorphs of formamidinium-based layered (110) hybrid perovskite

Sergey A. Fateev [1#], Andrey A. Petrov [1#], Ekaterina I. Marchenko [2], Yan V. Zubavichus [2], Victor N. Khrustalev [3,4], Andrey V. Petrov [5], Sergey M. Aksenov[6], Eugene A. Goodilin [1,7], Alexey B. Tarasov *[1,7]

[1]*Laboratory of New Materials for Solar Energetics, Faculty of Materials Science, Lomonosov Moscow State University; 1 Lenin Hills, 119991, Moscow, Russia;*

[2]*Federal Research Center Boreskov Institute of Catalysis, 5 Lavrentiev Ave., 630090, Novosibirsk, Russia;*

[3]*Inorganic Chemistry Department, Peoples' Friendship University of Russia (RUDN University), 6 Miklukho-Maklay Str., 117198, Moscow, Russia;*

[4]*N.D.Zelinsky Institute of Organic Chemistry RAS, 47 Leninsky Prosp., 119991, Moscow, Russia;*

[5]*Institute of Chemistry, Saint-Petersburg State University, 26 Universitetskii prospect, 198504 Saint-Petersburg, Russia;*

[6]*Laboratory of Nature-Inspired Technologies and Environmental Safety of the Arctic, Kola Science Centre, Russian Academy of Sciences, 14 Fersman str., Apatity, 184209, Russia;*

[7]*Department of Chemistry, Lomonosov Moscow State University; 1 Lenin Hills, 119991, Moscow, Russia;*

[#] *– these authors contributed equally*

*E-mail: alexey.bor.tarasov@yandex.ru*

# Section 1: Crystal structure of two polytypes of FA$_2$PbBr$_4$

Crystallographic data have been deposited with the Cambridge Crystallographic Data Center, CCDC 1967975. The supplementary crystallographic data can be obtained free of charge from the Cambridge Crystallographic Data Centre via www.ccdc.cam.ac.uk/data_request/cif.

The first FA$_2$PbBr$_4$ polytype of monoclinic crystal system was denoted as m-FA$_2$PbBr$_4$, and the second triclinic polytype as t-FA$_2$PbBr$_4$. The crystallographic and structural refinement parameters for two phases presented in Table S1.

As for the second polytype, we have made several attempts to refine its structure using different crystals. Despite the faceted shape and transparency of crystals (Figure S1), they all had a low internal quality, apparently due to the presence of large stacking disorder. This matches very well with so-called order/disorder phenomenon (related to polytypism and polymorphism) realized in a small tilting and rotating of adjacent octahedral layers around the axis which is perpendicular to the plane of the layer. In this case the diffraction patterns are characterized by a low quality with weak and diffuse peaks.

The "best" dataset for the t-FA$_2$PbBr$_4$ phase was independently processed twice with slightly different assumptions. The first refinement was performed using the whole set of diffraction peaks to reach a good completeness (0.994, Table S1). The hydrogens were removed from the refinement cycles. Accurate refinement of the model without fixing of any distances or angles gives us the final R ~ 16.20 based on 5529 I >3sI. This value of the final R1 is rather high, but the result is comparable with those for the double layered hydroxide (DLH)-structures or clays and mica structures with similar stacking disorder.

However, the resulting model shows crystal chemical reasonability and can be considered as an average structure. The searching of the symmetry relationship between the layers which describe the whole set of the polytypes will be the object of our further crystallographic investigations.

The explanations for all observed Alerts were added to the corresponding CIF. In particular, the twinning was checked using the proposed twin matrix. However, the refined twin domain volume is less than 3% and, therefore, it was ignored. The mistake with the crystal size was fixed. The standard multi-scan absorption correction was used.

For another refinement, we excluded the 287 low-angle reflections with inaccurately measured intensities from the final steps of the refinement due to some intensity overloads for this part of the detector. This minor fraction of intensity overloads was allowed in order to achieve better I/sigma statistics for high-angle reflections. However, this assumption resulted in rather low completeness (94.2%). In addition, internally low quality of crystal also led to some systematic intensity distortions which can be explained by the high level of stacking disorder.

**Table S1.** Crystallographic and structural refinement data for two polytypes of FA$_2$PbBr$_4$. The asterisk denotes the variant of refinement for t-FA$_2$PbBr$_4$ with excluded reflections.

| Compound | m-FA$_2$PbBr$_4$ | t-FA$_2$PbBr$_4$ | t-FA$_2$PbBr$_4$* |
|---|---|---|---|
| Empirical formula | C$_4$H$_{20}$Br$_8$N$_8$Pb$_2$ | C$_4$H$_{20}$Br$_8$N$_8$Pb$_2$ | C$_4$H$_{20}$Br$_8$N$_8$Pb$_2$ |
| $fw$ | 616.94 | 1213.71 | 1233.88 |
| $T$, K | 100 | 100 | 100 |
| Crystal size, mm | 0.17×0.16×0.04 | 0.20×0.04×0.04 | 0.20×0.04×0.04 |
| Crystal system | Monoclinic | Triclinic | Triclinic |
| Space group | $P\,2_1/c$ | $P$-1 | $P$-1 |
| $a$, Å | 13.1313(9) | 8.5642(17) | 8.5642(17) |
| $b$, Å | 7.9763(6) | 11.884(2) | 11.884(2) |
| $c$, Å | 11.7972(8) | 14.312(3) | 14.312(3) |
| $\alpha$, ° | 90 | 65.92(3) | 65.92(3) |
| $\beta$, ° | 91.4752(15) | 73.01(3) | 73.01(3) |
| $\gamma$, ° | 90 | 83.76(3) | 83.76(3) |
| $V$, Å$^3$ | 1235.22 | 1271.7(6) | 1271.7(6) |
| $Z$ | 4 | 2 | 2 |
| $d_c$, g cm$^{-3}$ | 3.318 | 3.17 | 3.222 |
| $F(0\,0\,0)$ | 1088 | 1048 | 1088 |
| $\mu$, mm$^{-1}$ | 26.556 | 32.517 | 32.519 |
| $\theta_{max}$, ° | 32.644 | 30.872 | 30.872 |
| Index range (h, k, l max) | 19, 12, 17 | 10, 15, 18 | 10, 15, 18 |
| No. of reflections collected | 4516 | 5564 | 5566 |
| No. of unique reflections ($R_{int}$) | 4490 (0.033) | 5529 (0.163) | 5242 (0.118) |
| No. of reflections with ($I > 2\sigma(I)$) | 3475 | 4446 | 4435 |
| Data / restraints / parameters | 4490 / 0 / 97 | 5529 / 0 / 140 | 5242 / 0 / 159 |
| $R_1$ ; w$R_2$ ($I > 2\sigma(I)$) | 0.0330; 0.0768 | 0.1620; 0.3456 | 0.1115; 0.2395 |
| $R_1$ ; w$R_2$ (all data) | 0.0332; 0.0785 | 0.1630; 0.3545 | 0.1188; 0.2434 |
| Goodness-of-fit (GOF) on $F^2$ | 1.034 | 2.500 | 1.065 |
| Extinction coefficient | 0.0023 | 0.0022(2) | 0.0022(2) |
| $T_{min}$; $T_{max}$ | 0.091; 0.816 | 0.060; 0.060 | 0.060; 0.060 |
| $\Delta\rho_{max}$; $\Delta\rho_{min}$, eÅ$^{-3}$ | 1.58; -2.02 | 20.87; -8.37 | 6.464; -4.798 |
| Data completeness | 99.4% | 99.4% | 94.2% |

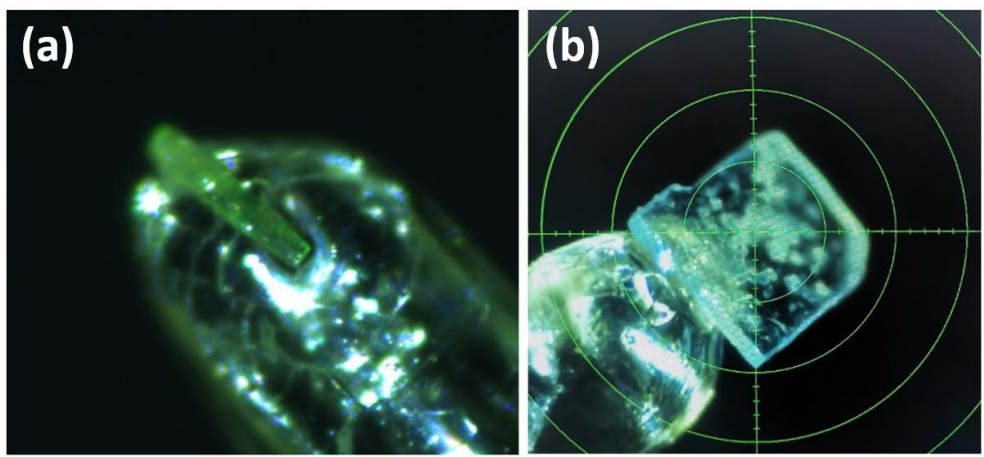

**Figure S1.** The appearance of t-FA$_2$PbBr$_4$ plate-like single crystals selected for refinement and mounted on sample holder tip. Opaque spots on the crystal surface are clots of Apiezon grease (was used to fix the crystal) and possibly droplets of residual solution.

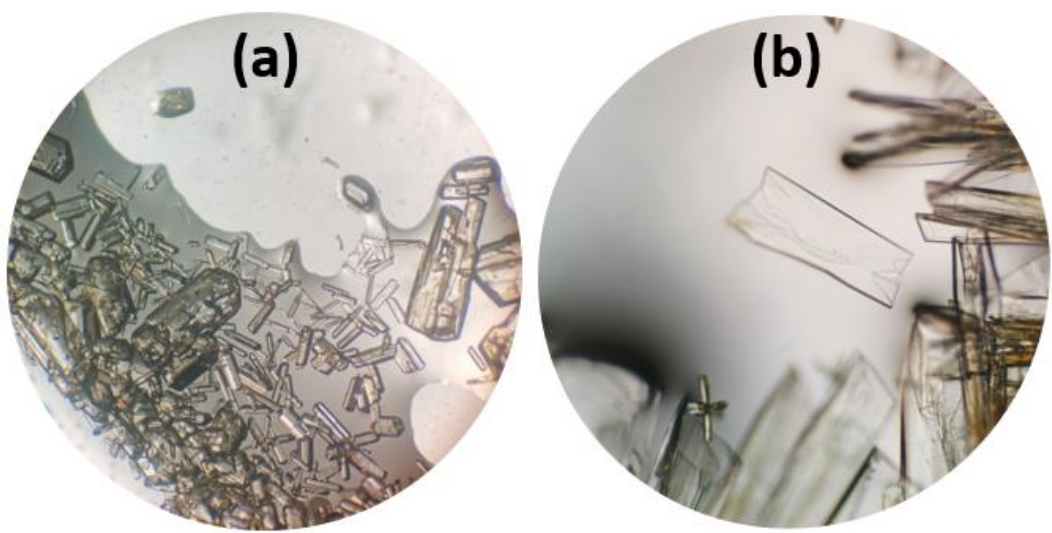

**Figure S2.** The crystallization from DMF solutions with a ratio FABr:PbBr$_2$ = 2.5:1. Photo (a) shows the case of fast evaporation resulting in faceted elongated prismatic crystals of the monoclinic phase (m-FA$_2$PbBr$_4$). The nucleation of this phase can occur both at the interface and in the bulk of the solution. Photo (b) shows the typical picture of triclinic (t-FA$_2$PbBr$_4$) phase crystallization, which usually forms under slow evaporation conditions, starts from the edge of droplet as a heterogeneous nucleation and grows further as thin transparent platelets.

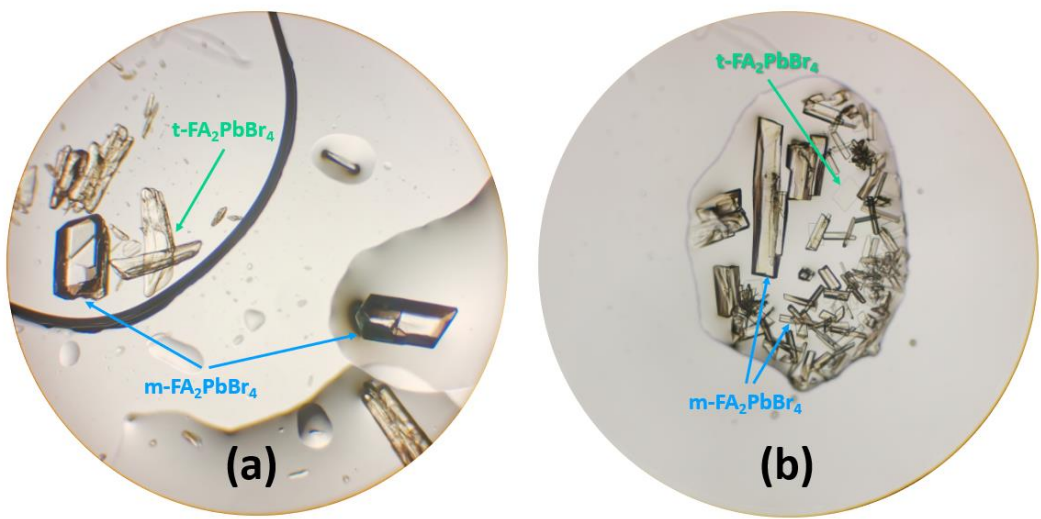

**Figure S3**. The examples of simultaneous crystallization of two polymorphs from a single droplet of solution (FABr:PbBr$_2$ = 2.5:1).

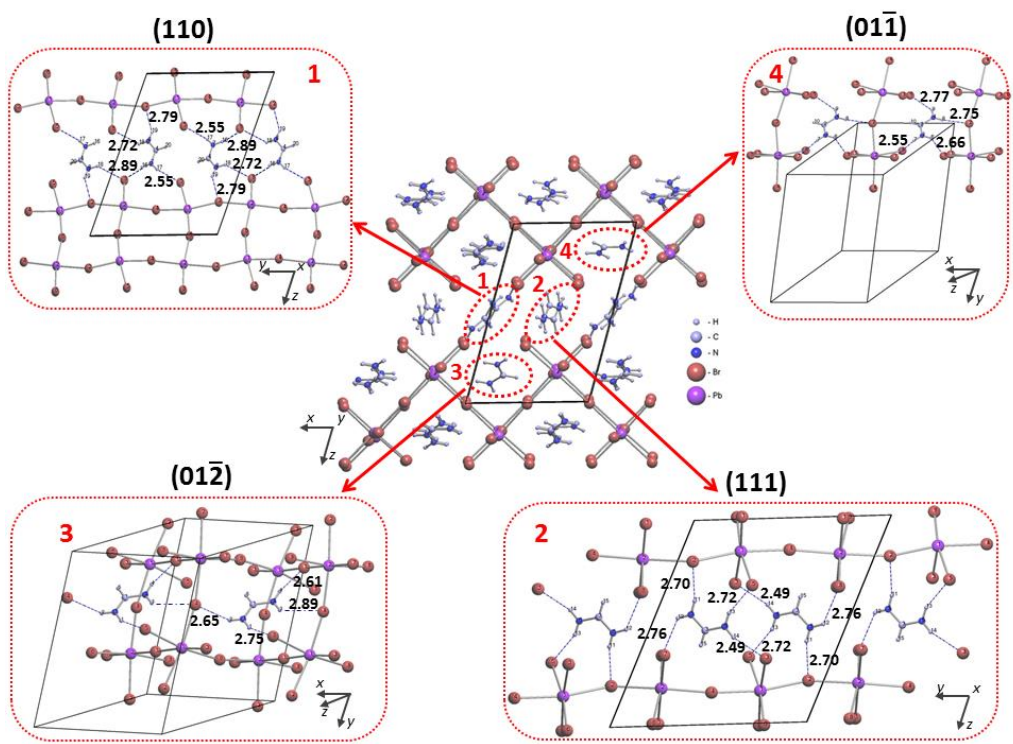

**Figure S4.** Crystal structure of t-FA$_2$PbBr$_4$ viewed down the b axis (in the center) and 4 projections on the planes (110), (111), (0-11), (01-2) containing flat FA$^+$ cations in 4 corresponding independent orientations.

# Section 2: Details of synthesis and phase equilibrium

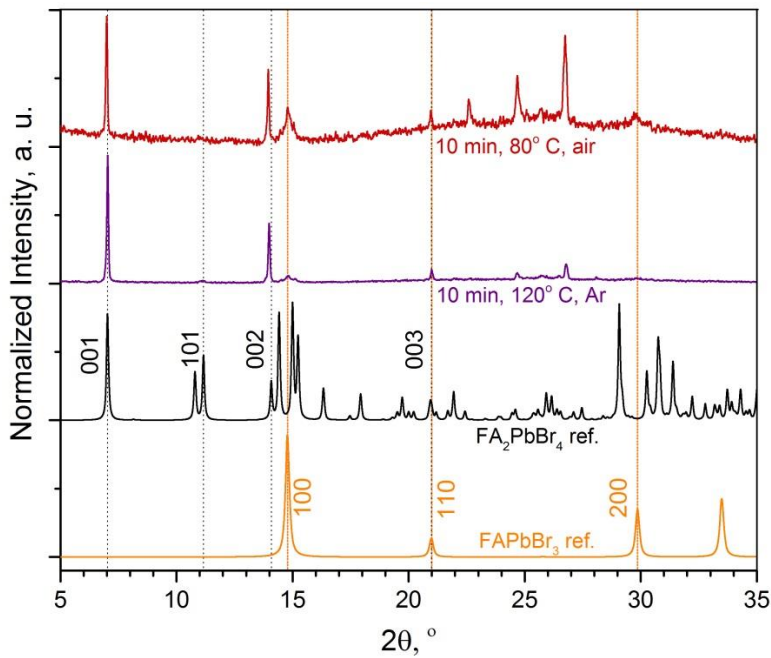

**Figure S5.** XRD pattern of the t-FA$_2$PbBr$_4$ film annealed during 10 min at 120 °C in argon glove box or in air (humidity ~ 35%) at 80 °C; at the bottom – two simulated powder diffraction patterns of t-FA$_2$PbBr$_4$ and FAPbBr$_3$ respectively.

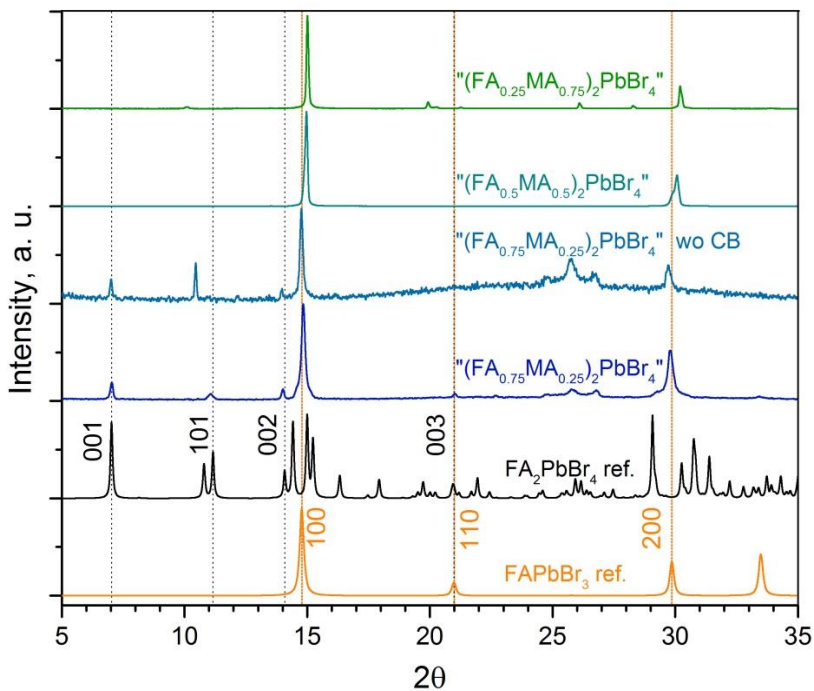

**Figure S6.** XRD pattern of the films of general composition A$_2$PbBr$_4$ (2ABr+PbBr$_2$) with different FABr/MABr ratio from 3:1 "(FA$_{0.75}$MA$_{0.25}$)$_2$PbBr$_4$" to 1:3 "(FA$_{0.25}$MA$_{0.75}$)$_2$PbBr$_4$" spin-coated with chlorobenzene dripping (unless otherwise indicated) at 25 s of rotation; at the bottom – two simulated powder diffraction patterns of t-FA$_2$PbBr$_4$ and FAPbBr$_3$ respectively.

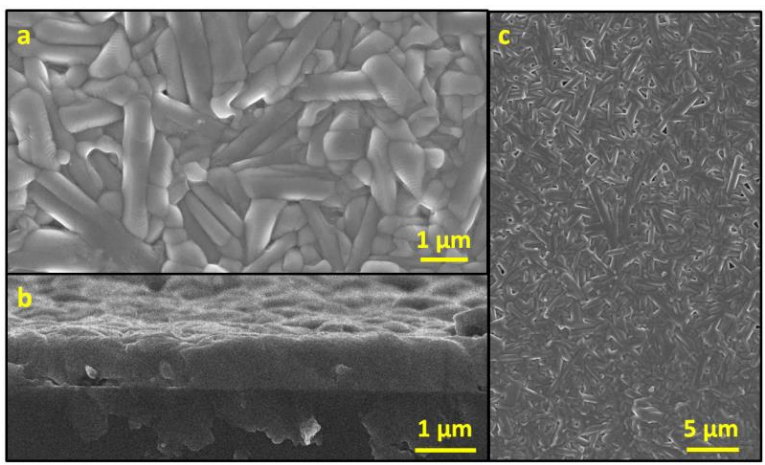

**Figure S7.** SEM micrographs of top surface (a, c) and cross-section (b) of the m-FA$_2$PbBr$_4$ thin film.

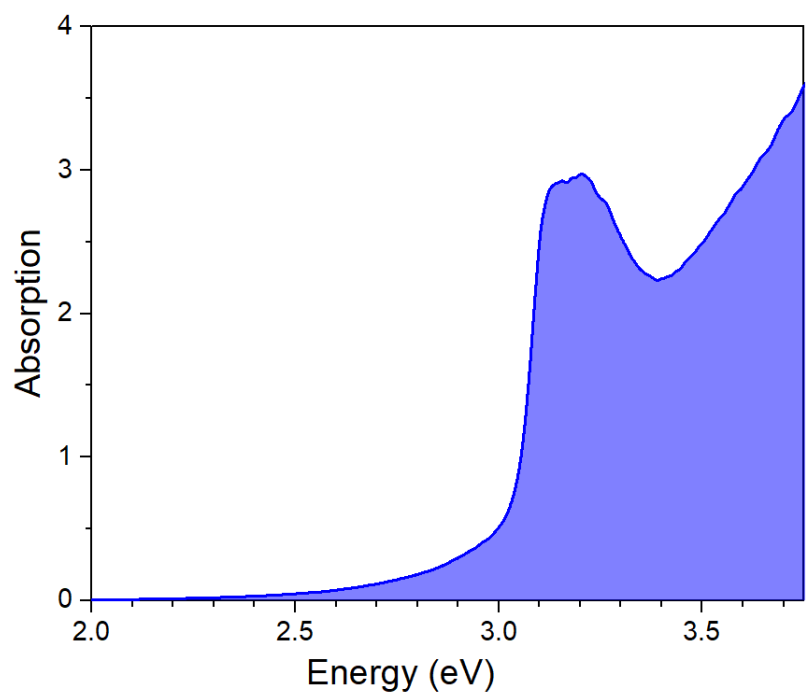

**Figure S8.** Absorption spectrum of the m-FA$_2$PbBr$_4$ thin film.

## Section 3: Details of the bandgap calculations

Assaying 3 different functionals (Table S2), we found that the differences in the shape of the bands are negligible, whereas the most plausible $E_g$ value (2.89 eV) was obtained from the HTCH, so it was used for all further calculations. All calculations were performed with atomic wave functions determined as a double numerical basis set with polarization (DNP) which is similar to 6-31G**[1].

**Table S2.** Bandgap values of t-$FA_2PbBr_4$ calculated using various functionals.

| Functional | PBE | PW91 | HCTH |
|---|---|---|---|
| $E_g$ (eV) | 2.88 | 2.86 | 3.14 |

*DFT calculations for model ($FA_2PbBr_4$-derived) structures*

Firstly, we showed that despite the strong hydrogen bonding the formamidinium cations have a negligible influence on the bandgap value of $FA_2PbBr_4$. To confirm this, we calculated the electronic structure for $FA_2PbBr_4$ unit cell without $FA^+$ cations and obtained almost similar value of the bandgap ($E_g$ = 3.14 eV with $FA^+$ and 3.11 eV without $FA^+$).

Then we performed calculations of the electronic structure for two series of model structures derived from real t-$FA_2PbBr_4$ which has 2 different interlayer Br···Br distance (d) of 4.15 Å and 4.63 Å. Since the structure contains the same number of distances of type 1 (4.15 Å) and type 2 (4.63 Å), the mean distance ($d_{mean}$) is 4.39 Å. The structures of the first series preserve the geometry of inorganic framework of the real $FA_2PbBr_4$ structure but have different $d_{mean}$ from 3.16 to 9.65 Å (Table S3, Fig.S4). The structures of the second series contain a inorganic framework similar to those of real $FA_2PbBr_4$ but without distortion of octahedral and without tilting (all Pb-Br bonds have the equal length and all Br-Pb-Br and Pb-Br-Pb angles are equal to 90° and 180° respectively). Because the second series have undistorted inorganic framework, all these structures possess only one interlayer distance (d).

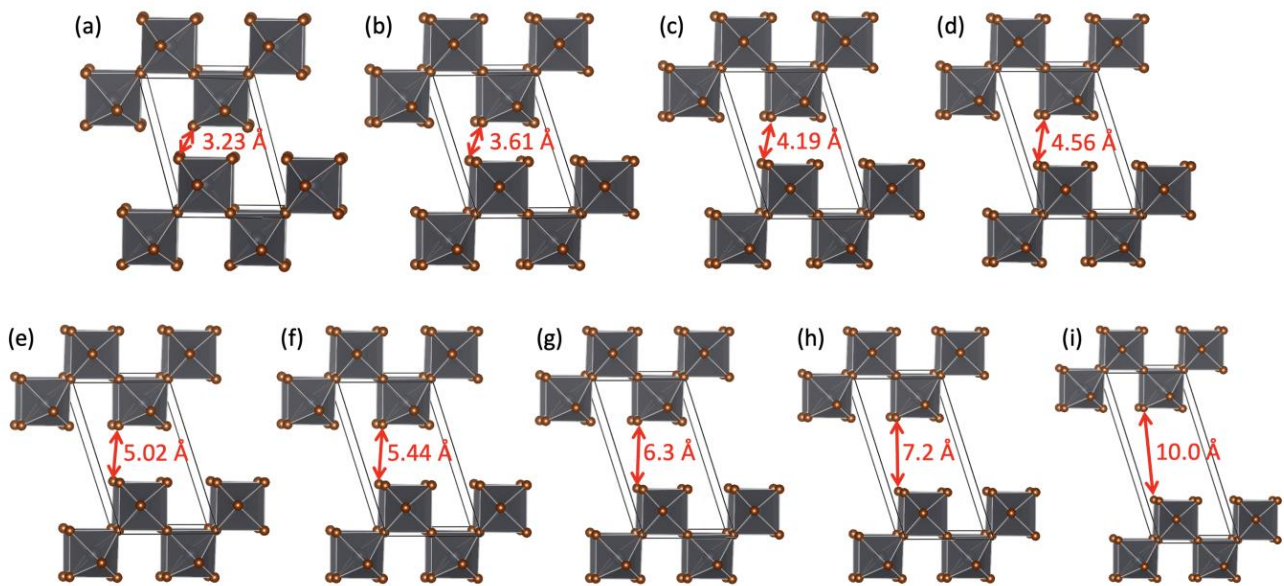

**Figure S9.** Model undistorted eclipsed structures with different interlayer distances (d) from 3.5 Å to 10 Å.

**Table S3.** The mean interlayer distances and calculated band gaps ($E_g$) of corresponding distorted (with inorganic framework identical to real $FA_2PbBr_4$) eclipsed model structures.

| Structure | $d_{mean}(Br \cdots Br)$, Å | $E_g$, eV |
|---|---|---|
| Structure 1 | 3.160 | 2.940 |
| Structure 2 | 3.465 | 3.011 |
| Structure 3 | 3.985 | 3.080 |
| Structure 4 | 4.320 | 3.106 |
| Structure 5 | 4.760 | 3.127 |
| Structure 6 | 5.165 | 3.140 |
| Structure 7 | 6.00 | 3.149 |
| Structure 8 | 6.88 | 3.153 |
| Structure 9 | 9.65 | 3.154 |

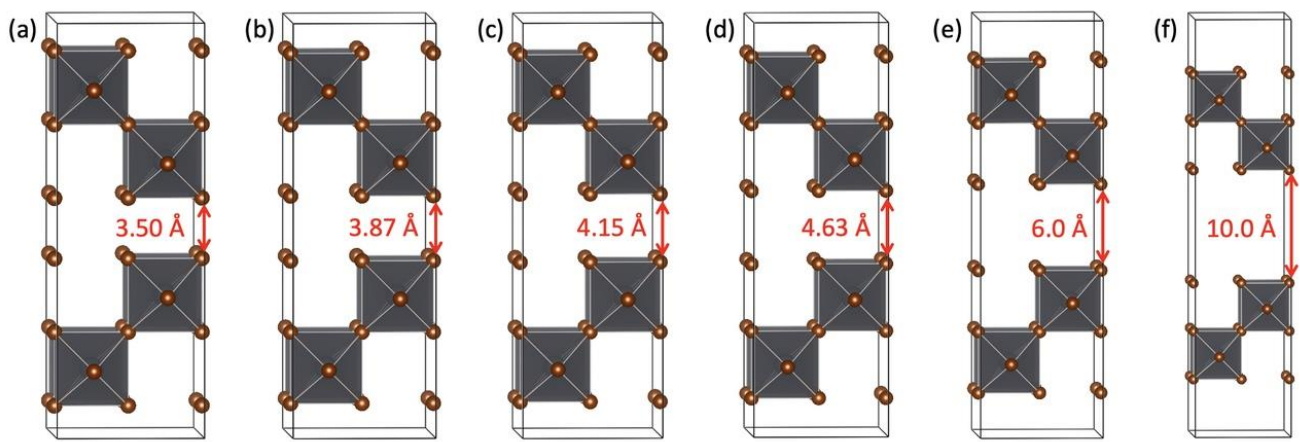

**Figure S10.** Model undistorted eclipsed structures with different interlayer distances (d) from 3.5 Å to 10 Å.

**Table S4.** The interlayer distances and calculated band gaps ($E_g$) of corresponding undistorted eclipsed model structures.

| Structure | $d(Br \cdots Br)$, Å | $E_g$, eV |
|---|---|---|
| Structure 1 | 3.50 | 2.866 |
| Structure 2 | 3.87 | 2.945 |
| Structure 3 | 4.15 | 2.978 |
| Structure 4 | 4.63 | 3.010 |
| Structure 5 | 6.00 | 3.037 |
| Structure 6 | 10.0 | 3.042 |

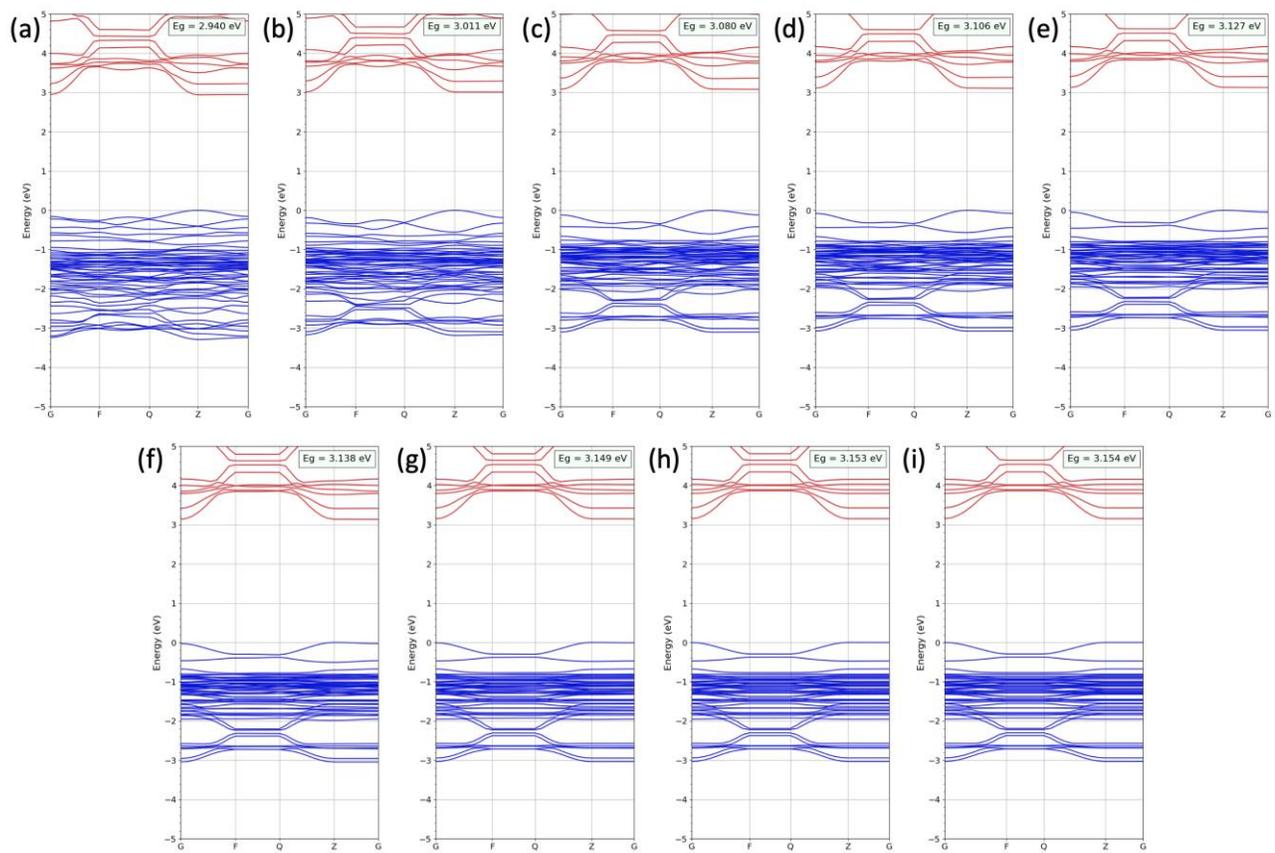

**Figure S11.** Calculated band structures of the model distorted eclipsed crystal structures of the FA$_2$PbBr$_4$ with different interlayer distances (d) from 3.5 Å to 10 Å.

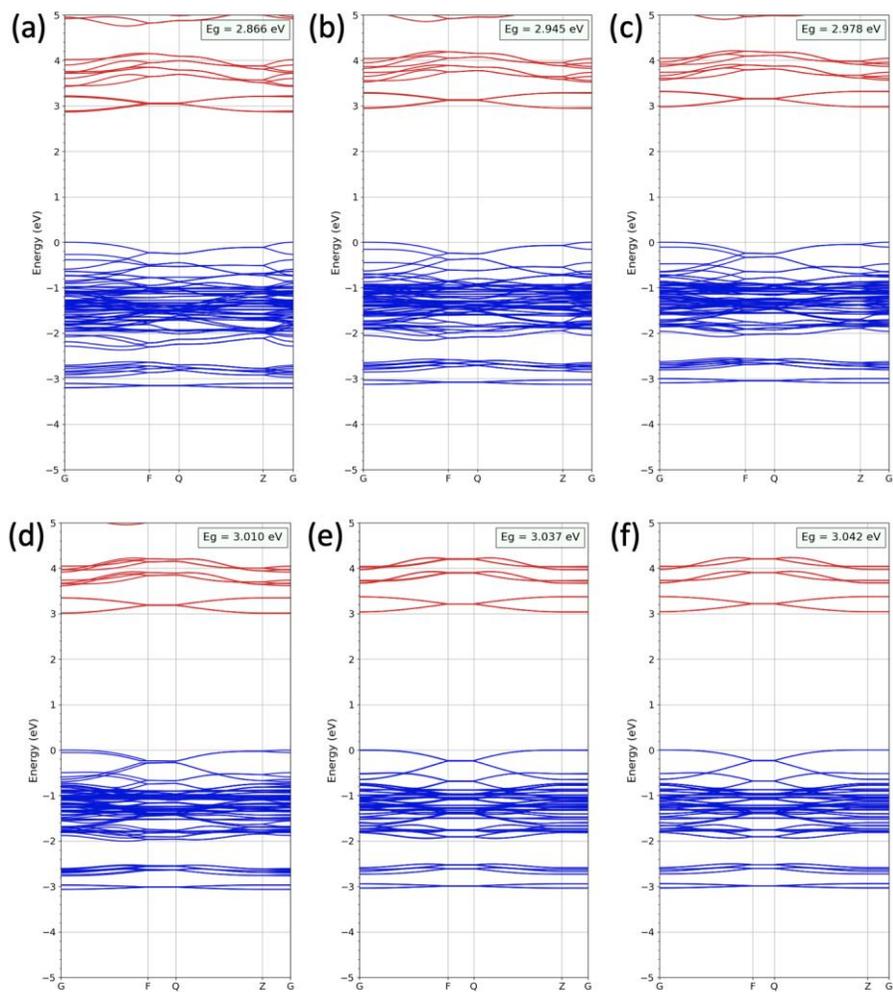

**Figure S12.** Calculated band structures of the model undistorted eclipsed crystal structures the FA$_2$PbBr$_4$ with different interlayer distances (d) from 3.5 Å to 10 Å.

# Section 4: Comparison of the FA$_2$PbBr$_4$ with other bromoplumbate perovskites

*Crystal chemical analysis of the specific volumes of moieties*

We consider the unit cell volume ($V_{U.C.}$) of a certain structure with formula A$_m$PbBr$_{m+2}$ as the sum of 2 volumes:

1) the total volume of organic cations A$^+$ per unit cell – $V_{A^+}^{tot}$,

2) the total volume of inorganic framework per unit cell – $V_{PbX}^{tot}$,

The total volumes ($V_{A^+}^{tot}$, $V_{PbX}^{tot}$) are calculated as two sum of the volumes of the A$^+$ organic cations and Pb$^{2+}$ plus Br$^-$ anions in certain crystallographic position respectively, as following:

$$V_{A^+}^{tot} = k\sum_{i=1}^{Z} V_{A^+}^i = kzV_{A^+}^{av.},$$

$$V_{PbX}^{tot} = \sum_{i=1}^{Z} V_{Pb}^i + (m+2) * \sum_{i=1}^{Z} V_{Br}^i = \sum V_{PbX}^{av.},$$

where Z is a number of formula units per unit cell; $V_{A^+}^i$, $V_{PbX}^i$ are specific volumes of A$^+$, [PbX$_4$] respectively. Because $V_{A^+}^i$ usually slightly varies at different positions within one unit cell it is reasonable to use average specific volumes $V_{A^+}^{av.}$, $V_S^{av.}$, $V_{PbX}^{av.}$. The fragmentation of the unit cell volume into "partial" specific volumes is performed using Voronoi-Dirichlet polyhedra (see details below), therefore, as a result, the sum of the volumes is equal to the unit cell volume ($V_{A^+}^{tot} + V_{PbX}^{tot} + V_S^{tot} = V_{U.C.}$).

*Details of the volumes calculation by Voronoi-Dirichlet polyhedra*

The unit cell of the FA$_2$PbBr$_4$ crystal structure contains 8 cations of formamidinium (half of these cations are symmetrically nonequivalent due to the fact that the structure was refined in the P$\bar{1}$ space group with Z=4). The average volume of the Voronoi-Dirichlet (V$_{VDP}$) polyhedron per one FA$^+$ cation is 96.79 Å$^3$ or 774.31 Å$^3$ per unit cell (see Figure S8 and Table S5), while the V$_{VDP}$ of the inorganic framework is 497.43 Å$^3$ per unit cell.

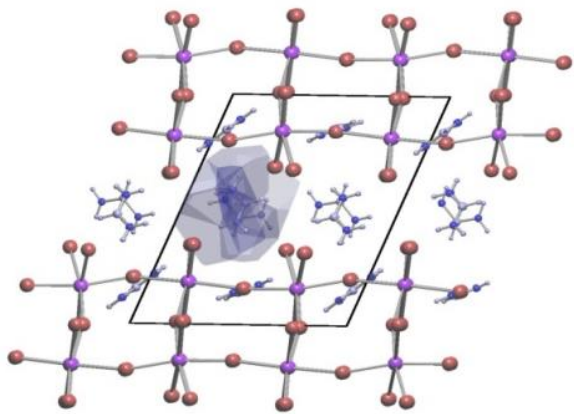

**Figure S13.** The inner volume of Voronoi-Dirichlet polyhedron of FA$^+$ cation in FA$_2$PbBr$_4$ crystal structure.

**Table S5.** The volumes of Voronoi-Dirichlet polyhedra for $FA^+$ cation containing in $FA_2PbBr_4$ unit cell.

| № of the $FA^+$ cation | V, Å$^3$ |
|---|---|
| 1 | 99.78 |
| 2 | 98.28 |
| 3 | 95.02 |
| 4 | 94.07 |
| 5 | 95.02 |
| 6 | 94.07 |
| 7 | 98.28 |
| 8 | 99.78 |

*Comparison of $FA^+$ and $[PbX_4]$ specific volumes in different crystal structures*

Using the average specific volumes of organic cations and inorganic PbX$_4$-moieties calculated from as-solved structure of t-FA$_2$PbBr$_4$, we can compare them with a corresponding specific volumes defined for related compounds described in the literature. In particular, it is reasonable to compare the volume of $FA^+$ cation in the structure of t-FA$_2$PbBr$_4$ with its volume in the structures of FAPbBr$_3$ and FAI, and compare the $V_{PbX}^{av.}$ for t-FA$_2$PbBr$_4$ with the same volume in the structures of related 110 bromide perovskites – (Im,Gua)PbBr$_4$ and (TzH,Gua)PbBr$_4$ in low-temperature $^{(lt)}$ and high-temperature $^{(ht)}$ modifications (Im, Gua and TzH is imidazolium, guanidinium and 1,2,4-triazolium cations respectively).

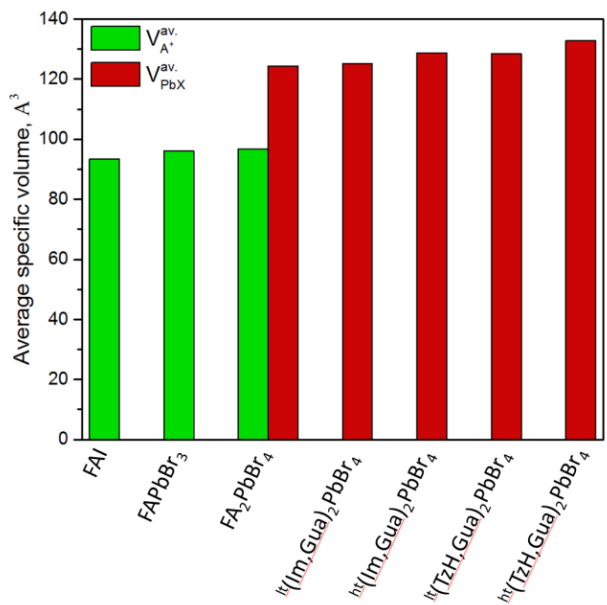

**Table S6.** The specific volumes of FA$^+$ and [PbX$_4$] in different crystal structures

| structure | $V_{FA^+}^{av.}$ | $V_{PbX}^{av.}$ |
|---|---|---|
| FAI | 93.37 | - |
| FAPbBr$_3$ | 96.13 | - |
| t-FA$_2$PbBr$_4$ | 96.79 | 124.4 |
| $^{lt}$(Im,Gua)PbBr$_4$ | - | 125.2 |
| $^{ht}$(Im,Gua)PbBr$_4$ | - | 128.8 |
| $^{lt}$(TzH,Gua)PbBr$_4$ | - | 128.5 |
| $^{ht}$(TzH,Gua)PbBr$_4$ | - | 132.8 |

**Figure S14.** The $V_{A^+}^{av.}$ and $V_{PbX4}^{av.}$ in different crystal structures

As one can see, the difference in the volumes of $FA^+$ cations in three structures does not exceed 2%, and in the case of FAPbBr$_3$ the difference is negligible. Similarly, the volume of the inorganic lattice for FA$_2$PbBr$_4$ is only slightly less than in similar 110 layered perovskites (taking into account that the measurement was carried out at 100 K it is more correct to compare with low-temperature modifications).

Based on this comparison, one can make a solid inference, that *the **structure of FA$_2$PbBr$_4$ is completely typical in the terms of specific volumes and density of packing point of view***.

**Table S7.** Comparison of the main geometrical parameters (bond angle variance $\sigma^2$ and bond length distortion Δd) of the FA$_2$PbBr$_4$ and similar (110) layered bromoplumbate perovskites.

| Composition | ID* | Min. interlayer distance, Å | $\sigma^2$ | Δd | FWHM of PL, meV | Ref |
|---|---|---|---|---|---|---|
| FA$_2$PbBr$_4$-t | 516 | 4.15 | 22.5 | 3.9 | 220 | This work |
| FA$_2$PbBr$_4$-m | 632 | 4.04 | 26.5 | 5.0 | - | This work |
| (N-MEDA)PbBr$_4$ | 58 | 4.39 | 24.5 | 8.2 | 880 | [2] |
| (Im,Gua)PbBr$_4$ | 203 | 4.49 | 22.1 | 7.6 | 90 | [3] |
| (Gua, Tz)PbBr$_4$ | 202 | 5.16 | 28.2 | 11.0 | 350 | [3] |
| (EDBE)PbBr$_4$ | 67 | 5.79 | 23.9 | 24.0 | 800 | [4] |
| (3-APy-dp)PbBr$_4$ | 373 | 4.32 | 21.1 | 9.8 | 743 | [5] |
| (3APIm-dp)PbBr$_4$ | 163 | 4.48 | 20.8 | 26.0 | >750 | [6] |
| (EPZ-dp)PbBr$_4$ | 375 | 3.90 | 38 | 41 | 370 | [7] |

* ID number in the database http://pdb.nmse-lab.ru/

Abbreviations for spacing organic cations:
N-MEDA = N-methylethane-1,2-diammonium [2];
Tz = $[C_2H_4N_3]^+$ = triazolium [3];
EDBE = $[NH_3(CH_2)_2O(CH_2)_2O(CH_2)_2NH_3]^{2+}$ = 2,2-(ethylenedioxy)-bis(ethylammonium) [4];
3APy-dp = $[(H_2N(CH_2)_3CH)NH_3]^{2+} = [C_4H_{10}N_2]^{2+}$ = diprotonated 3-aminopyrrolidine [5];
3APIm-dp = $[C_3H_4N_2(CH_2)_3NH_3]^{2+} = C_6H_{13}N_3$ = diprotonated N-(3-aminopropyl)imidazole [6];
EPZ-dp = $[(H_2NC_4H_8NH)CH_2CH_3]^{2+} = C_6H_{16}N_2$ = diprotonated 1-ethylpiperazine [7].

The emission band of FA$_2$PbBr$_4$ has a significantly narrower FWHM of 0.22 eV (52 nm) in comparison with a typical FWHM of bounded excitons induced photoluminescence (Table S7, Table S8). Among the (110) bromoplumbate layered perovskites known so far, only the (Im,Gua)PbBr$_4$ has a FWHM of PL emission (0.08 eV) smaller than FA$_2$PbBr$_4$. Both the compounds have an unusually small degree of PbX$_6$ octahedra distortion: Δd = 7.6·10$^{-4}$ for (Im,Gua)PbBr$_4$ and 3.9·10$^{-4}$ for FA$_2$PbBr$_4$ (Table S7).

Table S8. Comparison of the main optoelectronic parameters of $FA_2PbBr_4$ and similar (110) layered bromoplumbate perovskites: $PL_{max}$ stands for center of photoluminescence emission band; the optical bandgaps were determined in all cases as first absorption edge on UV-Vis spectrum; abbreviations FE and STE stands for free excitons and bounded excitons respectively.

| Composition | min. inter-layer distance, Å | $PL_{max}$, eV | FWHM of PL, meV | Optical bandgap, eV | Ref |
|---|---|---|---|---|---|
| $FA_2PbBr_4$(m) | 4.04 | - | - | 3.00 | This work |
| $FA_2PbBr_4$(t) | **4.15** | 2.91** | 105 | 2.82 | This work |
| | | 2.45*** | 235 | | |
| (N-MEDA)$PbBr_4$ | 4.39 | $2.22^{(STE)}$, $2.95^{(FE)}$ | 880 | 2.99 | [2] |
| (Im,Gua)$PbBr_4$ | 4.49 | 2.20 | 80 | 2.88 | [3] |
| (Gua,Tz)$PbBr_4$ | 5.16 | $1.77^{(STE)}$, $2.36^{(STE)}$ | 350 | 2.95 | [3] |
| (EDSTE)$PbBr_4$ | 5.79 | $2.17^{(STE)}$ | 800 | 3.22 | [4] |
| (3-APy-dp)$PbBr_4$ | 4.32 | $2.10^{(STE)}$ | 743 | 2.97 | [5] |
| (3APIm-dp)$PbBr_4$ | 4.47 | $2.10^{(STE)}$, $2.92^{(FE)}$ | >750 | 2.92 | [6] |
| (EPZ-dp)$PbBr_4$ | 3.90 | $2.08^{(ST)}$ | 370 | 3.12 | [7] |

\* - no intrinsic PL, ** - excitonic band, *** - unknown band probably related to surface defects

# References

[1] Inada Y., Orita H. Efficiency of numerical basis sets for predicting the binding energies of hydrogen bonded complexes: Evidence of small basis set superposition error compared to Gaussian basis sets // J. Comput. Chem. 2008. Vol. 29, № 2. P. 225–232.

[2] Dohner E.R., Hoke E.T., Karunadasa H.I. Self-Assembly of Broadband White-Light Emitters // J. Am. Chem. Soc. 2014. Vol. 136, № 5. P. 1718–1721.

[3] Guo Y.-Y. et al. Structure-directing effects in (110)-layered hybrid perovskites containing two distinct organic moieties // Chem. Commun. 2019. Vol. 55, № 67. P. 9935–9938.

[4] Dohner E.R. et al. Intrinsic White-Light Emission from Layered Hybrid Perovskites // J. Am. Chem. Soc. 2014. Vol. 136, № 38. P. 13154–13157.

[5] Li X. et al. Small Cyclic Diammonium Cation Templated (110)-Oriented 2D Halide (X = I, Br, Cl) Perovskites with White-Light Emission // Chem. Mater. 2019. Vol. 31, № 9. P. 3582–3590.

[6] Li Y.Y. et al. Novel 〈110〉-oriented organic-inorganic perovskite compound stabilized by N-(3-aminopropyl)imidazole with improved optical properties // Chem. Mater. 2006. Vol. 18, № 15. P. 3463–3469.

[7] Mao L. et al. Structural Diversity in White-Light-Emitting Hybrid Lead Bromide Perovskites // J. Am. Chem. Soc. 2018. Vol. 140, № 40. P. 13078–13088.

[8] Mączka M. et al. Layered Lead Iodide of [Methylhydrazinium]2PbI4 with a Reduced Band Gap: Thermochromic Luminescence and Switchable Dielectric Properties Triggered by Structural Phase Transitions // Chem. Mater. 2019. Vol. 31, № 20. P. 8563–8575.